\documentclass[onecolumn,showkeys,preprintnumbers,aps,a4paper,amssymb,prd,superscriptaddress,nofootinbib]{revtex4-2}
\usepackage{comment}
\usepackage{graphicx}
\usepackage{listings}
\usepackage{epsf}
\usepackage{bm}
\usepackage{amsmath}
\usepackage{amsfonts}
\usepackage{amssymb}
\usepackage{epstopdf}
\usepackage{color}
\usepackage[dvipsnames]{xcolor}
\usepackage{verbatim}
\usepackage{multirow}
\usepackage{soul}
\usepackage{physics}
\usepackage{bm}

\usepackage[width=0.00cm, height=0.00cm, left=1.50cm, right=1.50cm, top=2.00cm, bottom=2.00cm]{geometry}
\usepackage{microtype}
\usepackage{lmodern}

\usepackage[colorlinks = true,
            linkcolor = blue,
            urlcolor  = blue,
            citecolor = blue,
            anchorcolor = blue]{hyperref}
\usepackage[capitalize]{cleveref}
\usepackage[normalem]{ulem}
\usepackage{enumitem}
\usepackage{booktabs}

\usepackage{lipsum}

\makeatletter\let\expandableinput\@@input\makeatother

\begin{document}



\begin{center}
		\vspace{0.4cm} {\large{\bf Testing a Sign-Switch Cosmological Model with Curvature through Latest Planck 2018, DESI DR2 and PantheonPlus\&SH0ES Observational Data
}} \\
		\vspace{0.4cm}
		\normalsize{Archana Dixit $^1$, Manish Yadav$^2$, Anirudh Pradhan $^3$, M. S. Barak$^4$ }\\
		\vspace{5mm}
		\normalsize{$^{1}$ Department of Mathematics, Gurugram University, Gurugram, Haryana 122505, India }\\
		\normalsize{$^{2,4 }$Department of Mathematics, Indira Gandhi University, Meerpur, Haryana 122502, India}\\ 

        	\normalsize{$^{3 }$ Centre for Cosmology, Astrophysics and Space Science (CCASS), GLA University, Mathura, Uttar Pradesh, India}\\ 
		\vspace{2mm}
			$^1$Email address:archana.ibs.maths@gmail.com \\
		    $^2$Email address: manish.math.rs@igu.ac.in\\
                $^3$Email address: pradhan.anirudh@gmail.com\\
             $^4$Email address: ms$_{-}$barak@igu.ac.in\\
\end{center}

 
\maketitle
{\noindent {\bf Abstract.}
We investigate the spatial geometry of the Universe within the framework of a sign-switch dark energy scenario by extending the recently proposed $\Lambda_{\rm s}$CDM model to include a free curvature parameter $\Omega_k$.In this framework, the effective cosmological constant undergoes a transition from a negative to a positive value at a characteristic redshift $z_{\dagger}$. Using the latest Planck 2018 cosmic microwave background (CMB) data, DESI DR2 baryon acoustic oscillation (BAO) measurements, and the PantheonPlus\&SH0ES Type Ia supernova sample, we derive joint constraints on the spatial curvature parameter $\Omega_k$ and other cosmological parameters.
We find that Planck data  itself slightly favors a closed universe within both the 
$\Lambda_{\rm s}$CDM$+\Omega_k$ and $\Lambda$CDM$+\Omega_k$ frameworks, although spatial flatness remains well within the allowed uncertainties. When low-redshift probes were included, the curvature constraints were significantly tightened. In particular, the full Pk18+DR2+PP\&SH0ES dataset yields  $\Omega_k = 0.0001 \pm 0.0014$ for the $\Lambda_{\rm s}$CDM model, indicating a universe that is remarkably consistent with spatial flatness.
We further analyzed the correlations between $\Omega_k$, $H_0$, and $S_8$, finding that the inclusion of curvature and a sign-switch dark energy component helps stabilize cosmological parameter estimates while remaining compatible with current observational constraints. Model comparison using AIC and Bayesian evidence shows that the $\Lambda_{\rm s}$CDM model receives inconclusive/weak observational support relative to $\Lambda$CDM.

\smallskip 


\section{Introduction}

One of the central goals of cosmology is to determine whether the universe is spatially flat or possesses a nonzero curvature (open and closed). Any departure from flatness has profound implications for the cosmic fate and Inflation of the universe, underscoring the importance of constraining both the sign and amplitude of $\Omega_k$. This question has attracted considerable attention in recent years, especially following the results from the Planck 2018 cosmic microwave background (CMB) observations data \cite{ref1,ref2,ref3,ref4,ref4a,ref5}. CMB observations offer important constraints on the spatial curvature of the universe. The CMB (including polarization and temperature) data alone indicate a closed geometry with constraints $\Omega_k = -0.044^{+0.018}_{-0.015}$ \cite{ref1}. Joint analyses of Planck CMB lensing and BAO data within the $\Lambda$CDM model are consistent with the spatially nearly open universe ($\Omega_k = 0.0007 \pm 0.0019$)\cite{ref1}. However, the inflation generically predicts a spatially flat universe \cite{ref6}, making flatness an important observational signature. A nonflat geometry, if confirmed, would cast doubt on standard inflationary scenarios and may signal either hidden systematics or extensions to $\Lambda$CDM model. Simultaneously, multiple late-time cosmological studies have yielded results compatible with a mildly open universe, showing a modest discrepancy with flat or closed geometries\cite{ref7,ref8}.\\

A wide range of methods has been developed to measure and constrain the spatial curvature of the universe. In particular, the model-independent approach introduced by \cite{ref9} combines strong gravitational-lensing time delays with supernova distance measurements to infer the curvature parameter with a precision of  $\sim 6 \times 10^{-3}$. The role of curvature in cosmological tests of General Relativity was highlighted in \cite{ref10}, where it was shown that neglecting curvature can lead to artificial deviations from GR and significantly alter the allowed growth parameter space. However, as argued in \cite{ref11}, achieving high-precision curvature measurements remains challenging, since robust constraints often rely on assumptions about the dark energy evolution and the adoption of standard $\Lambda$CDM parameter values. An alternative strategy based on a test-particle framework was proposed in \cite{ref12}, demonstrating that curvature can be inferred using at least six particles in a general spacetime or four particles in vacuum. Model-independent determinations of cosmic curvature using gravitational-wave standard sirens and cosmic chronometers (CC) have also attracted considerable attention in recent numerical studies. The potential of these probes was demonstrated in \cite{ref13} and \cite{ref14}, with the latter suggesting that future DECIGO space-based gravitational-wave detector could deliver reliable and competitive curvature constraints. Further investigations employing gravitational waves and CC were carried out in \cite{ref15} and \cite{ref16}, where \cite{ref16} showed that combining gravitational-wave observations with strong gravitational-lensing systems allows for independent measurement of cosmic curvature. Other methodologies have also been proposed to address this problem. Among them, the distance sum rule (DSR) provides a purely geometric test, rooted in the triangle inequality of curved space, and is therefore largely model independent \cite{ref17,ref18,ref19,ref20,ref21,ref22,ref24}. In the FLRW universe, the angular comoving diameter distances connecting the source observer, and lens at three distinct redshifts obey a curvature-dependent relation that deviates from linearity when  $\Omega_k \neq 0$. Any statistically significant deviation from this relation either signals a breakdown of the FLRW metric or provides direct, model-independent constraints on the spatial geometry of the universe. Observational implementations of DSR have relied on reconstructing these distances using data from supernovae, quasars, and CC. Such reconstructions have been performed using both parametric and non-parametric techniques, including Gaussian processes and deep-learning methods \cite{ref25,ref26,ref27,ref28,ref29}. The current results indicate that curvature constraints inferred from  DSR are sensitive to assumptions about lens modelling, the adopted reconstruction methodology, and the quality of the underlying datasets.\\

In the most recent works on the spatial curvature of the universe using updated observational datasets, the author \cite{ref30} investigated a $w$CDM model with latest DESI data (DR1 and DR2) and other data such as CC, and PantheonPlus (PP), and reported indications of a non-zero cosmic curvature deviation with more than $1\sigma$ level. In the subsequent analysis, Ref. \cite{ref31} employed CPL parametrization from an extended data combination including CC+DR2+PP+$r_d$ +$f\sigma_8$ and found open geometry $10^2\Omega_k = 1.8$ of the universe. A similar result was obtained by Pengu \cite{ref32,ref33} using five different dynamical dark energy models. Several varieties have been presented in the literature based on the spatial curvature of the universe using different theories, methodologies, and combinations of datasets \cite{ref34,ref35,ref36,ref37,ref38,ref39,ref40,ref41,ref42,ref43,ref44}.\\

Motivated by recent advances in studies on cosmic curvature, we extend our previous investigation of sign-switch dark energy \cite{ref45} by incorporating the effects of spatial curvature. In this study, we confront the sign-switch dark energy model with a comprehensive combination of the latest DR2 data, CMB, and PP\&SH0ES sample. Our primary objective in this study was to constrain the curvature density parameter alongside other cosmological parameters and to examine how these constraints depend on different combinations of observational datasets. The remainder of this paper is organized as follows. Section I provides a comprehensive overview of recent developments in studies of the spatial curvature of the universe. In Section II, we introduce the sign-switch dark energy model considered in this study. Section III outlines the methodology and describes the observational datasets used the analysis. Section IV presents. The main results and their cosmological implications. Finally, Section V summarizes the findings and discusses their broad significance.\\


\section{MODEL}

In this work, we consider a recently proposed and promising extension of the standard $\Lambda$CDM model, namely $\Lambda_{\rm s}$CDM \cite{ref46,ref47,ref48,ref49,ref50,ref51}. This model emerges from insights developed within the graduated dark energy (gDE) scenario, where it has been conjectured that the universe may have experienced a smooth transition from an anti–de Sitter (AdS) vacua to de Sitter (dS) vacua during its late-time evolution. Motivated by this possibility, we consider a sign-switching cosmological constant \cite{ref45}, denoted by $\Lambda_{\rm s}$, which changes sign from negative (past)  to positive (present) at transition redshift $z_{\dagger}$, here, $z_{\dagger}$ treated as a free parameter to be constrained by observations. The transition redshift $z_{\dagger}$ reported in the literature is not well constrained by CMB data alone \cite{ref46}. However, when the full BAO dataset is combined with CMB and supernova observations, the transition is found to occur at $z_{\dagger} > 2.36$ at the $95\%$ confidence level \cite{ref44}. Upon including Ly-$\alpha$ data, the transition shifts to $z_{\dagger} \sim 1.84$ at the $68\%$ confidence level \cite{ref47}. Moreover, in the absence of BAO data, the transition is also reported to lie around $z_{\dagger} \sim 1.78$ \cite{ref48,ref50}. These results indicate that the value of $z_{\dagger}$ is strongly dependent on the choice of observational datasets, although a mild correlation with cosmological parameters is also observed. In the $\Lambda_{\rm s}$CDM model, the constant vacuum energy density of the standard $\Lambda$CDM paradigm is replaced by its dynamic sign-switching counterpart.

Formally, the behaviour of $\Lambda_{\rm s}$ is encoded through the signum function as
\begin{equation}
	\label{eq1}
\Lambda \ \longrightarrow\ \Lambda_{\rm s} \equiv
\Lambda_{\rm s0} \  {\rm sgn}(z_\dagger - z),
\end{equation}
where $\Lambda_{\rm s0} > 0$ corresponds to the present value of the cosmological term, $z_\dagger$ denotes the redshift at which the sign transition occurs, and ${\rm sgn}(x)$ is the usual signum function. Thus, for $z > z_\dagger$ (the early universe), the model predicts an AdS-like phase with $\Lambda_{\rm s} < 0$, whereas for $z < z_\dagger$, the vacuum energy becomes positive, leading to the observed late-time acceleration.

The cosmological dynamics of the $\Lambda_{\rm s}$CDM scenario are governed by a modified Friedmann equation with the geometry of the universe.
\begin{equation}\label{eq2}
\frac{H^2(z)}{H_0^2}
= \Omega_{r0}(1+z)^4 + \Omega_{m0}(1+z)^3 + \Omega_{k0}(1+z)^{2} + \Omega_{\Lambda_{s0}}{\rm sgn}(z_\dagger - z),
\end{equation}
where $\Omega_{r0}$, $\Omega_{m0}$, and $\Omega_{\Lambda_{s0}}$ are the current density parameters of radiation, matter, and  sign-switching vacuum component, respectively. \\
The curvature density parameter $\Omega_{k}$ is related to the spatial curvature $k$ of the universe as follows:
\begin{equation}
    \Omega_{k0} = -\frac{c^{2}k}{3H_{0}^{2}},
\end{equation}
where $c$ denotes the speed of light in vacuum. The sign of the curvature parameter $k$ determines the global geometry of the universe:
\begin{itemize}
    \item \(k < 0\):  \textit{open} (negatively curved) universe,
    \item \(k > 0\):  \textit{closed} (positively curved) universe,
    \item \(k = 0\):  \textit{spatially flat} universe.
\end{itemize}

In this study, we introduce our sign-switch model with curvature ($\Lambda_{\rm s}$CDM+$\Omega_k$) ,

\begin{equation}\label{eq4}
H(z)
= H_0 \sqrt {\Omega_{r0}(1+z)^4 +\Omega_{m0}(1+z)^3 + \Omega_{{k}0}(1+z)^{2} + \Omega_{\Lambda_{\rm s0}}{\rm sgn}(z_\dagger - z)},
\end{equation}

All the  density parameters present in our sign switch model satisfy the  nearly condition:  $ \Omega_{r0}+\Omega_{m0}+  \Omega_{k0}+ \Omega_{\Lambda_{s}0} = 1$\\

 A sign-changing dark energy term offers a distinctive phenomenological framework capable of addressing certain tensions in modern cosmology—most notably, those related to the expansion history and potential deviations from the standard $\Lambda$CDM behavior at redshifts $z \sim 1$–$3$. The $\Lambda_{\rm s}$CDM+$\Omega_k$ model thereby provides a well-defined and theoretically motivated avenue for exploring possible extensions to the concordance model while remaining sufficiently simple to be confronted with current and future observational datasets.\\
 
 The deviations of the $\Lambda_s$CDM model from the standard $\Lambda$CDM scenario are governed by the additional parameter $z_\dagger$, which determines the redshift at which the cosmological constant undergoes a sign transition. For redshifts $z < z_\dagger$, the two models are effectively identical, as they share the same present-day matter density $(\Omega_{m0}h^2)$ and cosmological constant ($\Lambda_{\rm s0} = \Lambda$). At redshifts $z > z_\dagger$, the models differ due to the sign reversal, $\Lambda_{\rm s} = -\Lambda_{\rm s0}$; however, this distinction becomes negligible at sufficiently high redshifts, where the corresponding density parameters $\Omega_{\Lambda_{\rm s}} = \Lambda_{\rm s}/(3H^2)$ and $\Omega_\Lambda = \Lambda/(3H^2)$ rapidly decay relative to the matter component. Consequently, the $\Lambda_{\rm s}$CDM model effectively departs from $\Lambda$CDM only within the intermediate redshift range $z_\dagger < z \ll z_*$, and can therefore be interpreted as a post-recombination modification of the standard cosmological model. An important consequence of this framework is that the inferred value of $H_0$ in the $\Lambda_{\rm s}$CDM model is generally higher than in $\Lambda$CDM, with $z_\dagger$ exhibiting an inverse correlation with $H_0$. Such higher values arise as a direct consequence of the sudden drop in $H(z)$ due to the negative cosmological constant for $z > z_\dagger$. Moreover, as $z_\dagger$ increases, $H_0$ decreases, approaching the $\Lambda$CDM value in the limit $z_\dagger \to \infty$. This behavior can be understood for two reasons: first, as $z_\dagger$ increases, the portion of the distance modulus (DM) integral over which $\Lambda_{\rm s}$ is negative decreases, thereby requiring less compensation from the positive $\Lambda_s$ contribution, including $H_0$; second, as $z_\dagger$ increases, the sign-switching feature of $\Lambda_{\rm s}$ becomes less effective, since for large $z_\dagger$ the Universe is already matter-dominated at the time of the transition, making the effect of negative $\Lambda_{\rm s}$ on the evolution of $H(z)$ negligible. For instance, at $z_\dagger = 3$, just before the cosmological constant becomes negative ($z \to z_\dagger^{-}$), the matter component already dominates, with $\Omega_m(z = 3) \approx 0.96$, corresponding to only $|\Omega_{\Lambda_{\rm s}}/\Omega_m| \approx 0.04$.\\
 
 
 \section{DATASETS AND METHODOLOGY}
\label{sec:datasets}

\textbf{Planck 2018 (Pk18)}: 

In this study, we adopted a highly reliable Pk18 legacy  Cosmic Microwave Background (CMB) dataset. It includes detailed measurements of the polarization power spectra  and temperature anisotropies, lensing reconstructions and  their cross-spectra,\cite{ref52,ref1},viz., the high-$\ell$ \texttt{Plik} likelihood for the ($30 \leq \ell \leq 1996$) TE and EE  as well as  for ($30 \leq \ell \leq 2508$) TT, the low-$\ell$  ($2 \leq \ell \leq 29$) EE-only likelihood based on the \texttt{SimAll} method, as well as measurements of the CMB lensing, the low-$\ell$  ($2 \leq \ell \leq 29$) EE-only TT-only likelihood  from \texttt{Commander} component-separation algorithm.\\

\textbf{Dark Energy Spectroscopic Instrument Data Release 2 (DR2)}: 

We use the most up-to-date BAO measurements from DESI DR2 \cite{ref53,ref53a}, including LRG1–2, BGS, LRG3+ELG1, QSO, ELG2, and Ly$\alpha$ observations samples over the range $0.295 < z < 2.33$ to constrain three comoving distances such as $D_{M}(z)$ , $D_{H}(z)$ , and $D_{V}(z)$, which are defined as, 

\begin{itemize}
\item  \textbf{Hubble distance}

$D_{H}(z)/r_{\rm d} = \frac{c}{H(z).r_{\rm d}}$

\item  \textbf{Transverse comoving distance}

$D_{M}(z)= S_{k}(x) \int_0^z \text{d}z' {c \over H(z')}$. Where $c$ is the speed of light and \\

$S_{k}(x) =
\begin{cases}
\dfrac{\sin\!\left(\sqrt{-\Omega_{k}}\,x\right)}{\sqrt{-\Omega_{k}}}, & \Omega_{k} < 0, \\[10pt]
x, & \Omega_{k} = 0, \\[10pt]
\dfrac{\sinh\!\left(\sqrt{\Omega_{k}}\,x\right)}{\sqrt{\Omega_{k}}}, & \Omega_{k} > 0.
\end{cases}$

\item \textbf{Angle-averaged distance}

$D_{V}(z)/r_{\rm d} \equiv \frac{\left[z D^2_M(z) D_H(z)\right]^{1/3}}{r_{\rm d}}$ 

\end{itemize}

where, $r_{\rm d} = \int_{z_{\rm d}}^{\infty} \frac{c_{\rm s}\, \mathrm{d}z}{H(z)}$ denotes the comoving sound horizon at the drag epoch $z_{\rm d}$, and $c_{\rm s}$ is the speed of sound of the baryon-photon fluid. The DR2 measurements used in our analysis are summarized in Table IV in Ref. \cite{ref53}.\\

\textbf{Pantheon Plus and SH0ES (PP\&SH0ES)}: In this study, we make use of the PP\&SH0ES dataset by incorporating the latest SH0ES Cepheid host distance calibrations~\cite{ref55} into the likelihood analysis, together with the distance modulus measurements from Type Ia supernovae (SNe Ia) in the Pantheon Plus compilation~\cite{ref54}. The Pantheon Plus dataset comprises 1701 light curves corresponding to 1550 distinct SNe Ia, covering a redshift range of $z \in [0.001,\,2.26]$.

The observed distance modulus is defined as
\begin{equation}
	\mu_{\mathrm{obs}} = m_B - M_B,
\end{equation}
where $m_B$ is the apparent magnitude and $M_B$ is the absolute magnitude of Type Ia supernovae. In this work, the absolute magnitude $M_B$ is effectively fixed by the SH0ES calibration and is not treated as an independent free parameter.\\

The theoretical distance modulus is given by
\begin{equation}
	\mu_{\mathrm{th}}(z) = 25 + 5 \log_{10} d_L(z),
\end{equation}
where the luminosity distance $d_L(z)$ with non-flat is defined as
\begin{equation}
	\label{eq:dl}
	d_L(z) = (1+z)\,S_k(x)\int_0^{z}\frac{dz^\prime}{H(z^\prime)}.
\end{equation}

The difference between $\mu_{\mathrm{obs}}$ and $\mu_{\mathrm{th}}$ is used to constrain the cosmological parameters. The full covariance matrix associated with the supernova sample is included in the likelihood analysis.\\

In this work, we employ the Boltzmann  CLASS  code \cite{ref56} together with the Bayesian analysis framework \texttt{MontePython} \cite{ref57} to derive cosmological constraints using the Markov Chain Monte Carlo (MCMC) technique. We aim to explore the sign switch dark energy model ($\Lambda_{\rm s}$CDM+$\Omega_k$). The  MCMC chain results were analyzed within GetDist \cite{ref58}, and the convergence of the chains was ensured by applying the Gelman–Rubin criterion \cite{ref59}, requiring $R - 1 \leq 0.01$. For each parameter constraint, we run four independent chains with 100,000 steps, discarding the first 30,000 steps as burn-it. We adopted uniform priors on the cosmological parameters as follows: $\Omega_b h^2 \in [0.005, 0.1]$, $\Omega_c h^2 \in [0.001, 0.99]$, $100\,\theta_{s} \in [1.03,\,1.05]$, $\ln(10^{10}A_{s}) \in [3.0,\,3.18]$, $n_{s} \in [0.9,\,1.1]$, and $\tau_{\rm reio} \in [0.04,\,0.125]$, $\Omega_k \in [-1,1]$  and $z_{\dagger} \in [1,\,3]$ for  $\Lambda_{\rm s}$CDM+$\Omega_{k}$ model.


\section{Results and discussion}\label{sec:results}

In this article, we first present the cosmological constraints obtained using the geometry of the universe ( free spatial curvature $\Omega_k$ parameter). Table \ref{tab1} summarizes the marginalized constraints on the cosmological parameters derived from Pk18 data alone, as well as from joint analyses combining Pk18 with DR2 and with the full Pk18+DR2+PP\&SH0ES datasets. To illustrate the impact of the different data combinations and model extensions, Fig. \ref{fig1} shows the two-dimensional marginalized posterior distributions in the $\Omega_k - H_0$ plane for both the $\Lambda_{\rm s}$CDM+$\Omega_k$ and $\Lambda$CDM+$\Omega_k$ model analyze with low-redshift and high-redshift given data combinations such as Pk18, Pk18+DR2, Pk18+DR2+PP\&SH0ES. The Pk18 alone provides an independent and highly precise probe for the spatial curvature of the universe. From the Pk18 data alone, we obtain a constraint of $\Omega_k = -0.0059^{+0.0064}_{-0.0054}$ within the $\Lambda_{\rm s}$CDM$+\Omega_k$ framework. This result is consistent with a mildly closed universe and indicates  a deviation from exact spatial flatness at approximately the $1.0\sigma$ level. For the $\Lambda$CDM$+\Omega_k$ model, the corresponding constraint is  $\Omega_k = -0.0095^{+0.0064}_{-0.0056}$, which similarly favors a closed geometry, with a slightly higher  statistical preference of approximately $1.6\sigma$. These findings are broadly consistent with the Planck 2018  CMB-only curvature constraints, which also show a mild preference for a closed universe while remaining  compatible with spatial flatness within the quoted uncertainties.\\

\begin{figure}[hbt!]
    \centering
    \includegraphics[width=0.48\linewidth]{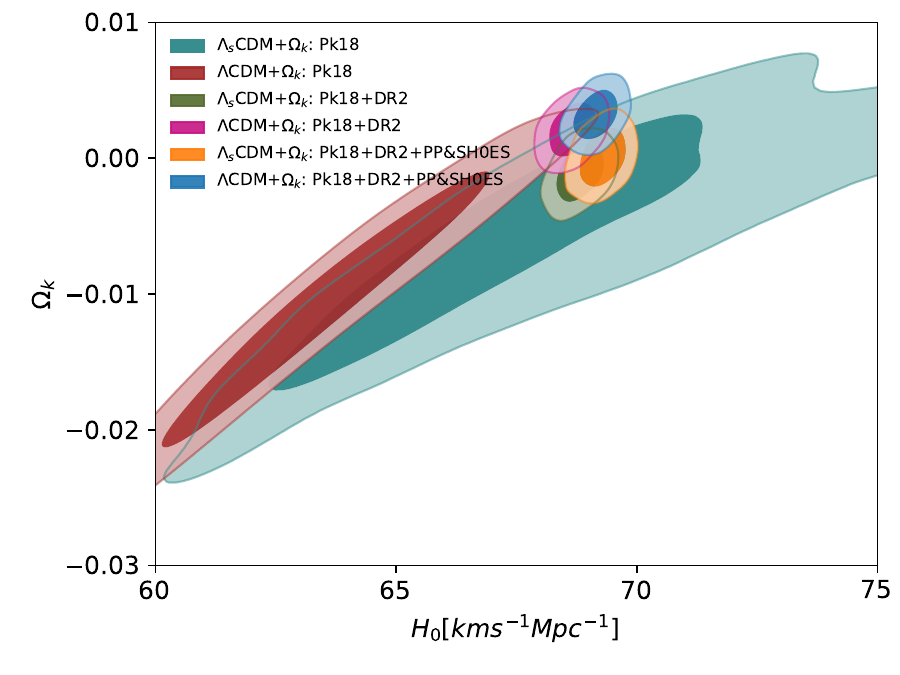}
    \includegraphics[width=0.48\linewidth]{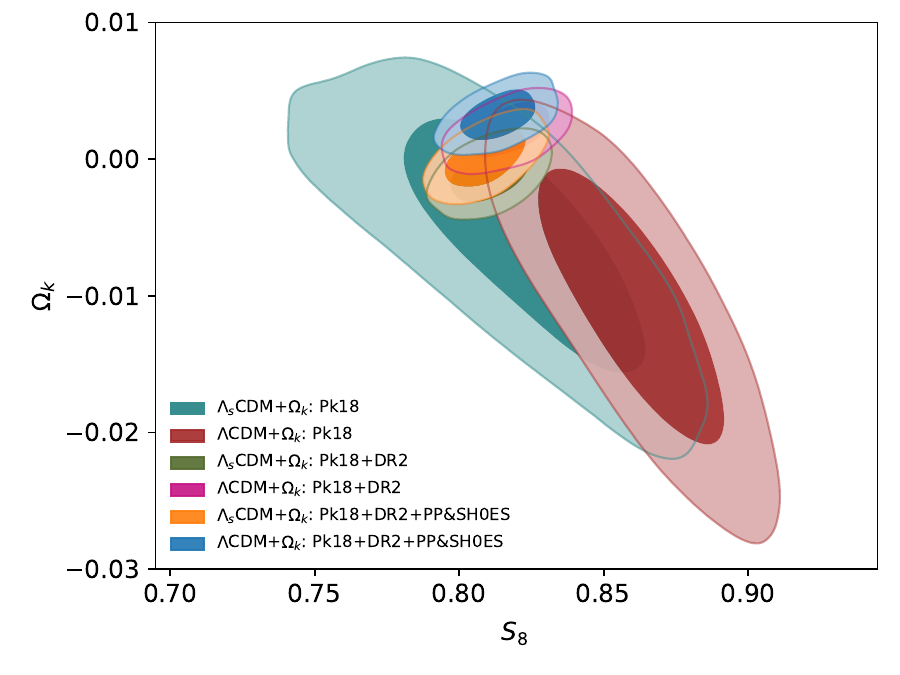}
    \caption{The 2D confidence contours (68\% and 95\% confidence levels) in the 
$\Omega_k$--$H_0$ plane (left panel) and the  $\Omega_k$--$S_8$ plane (right panel) for the 
$\Lambda_{\rm s}$CDM and $\Lambda$CDM models, obtained from all the considered data combinations.}
    \label{fig1}
\end{figure}

Next, we combined the Pk18 data with DR2 measurements to further tighten constraints on the spatial curvature (geometry) of the universe. As shown in both panels of  Fig.\ref{fig1}, the inclusion of DR2 data significantly reduced the uncertainty of $\Omega_k$ relative to the Pk18-only analysis. Within the $\Lambda_{\rm s}$CDM$+\Omega_k$ framework, the joint Pk18+DR2 analysis yielded $\Omega_k = -0.0010 \pm 0.0014$, corresponding to a nearly closed geometry of the present universe. The preferred curvature in this model is slightly shifted toward negative values, although the deviation from the spatial flatness remains below the $1\sigma$ level. In contrast, for the $\Lambda$CDM$+\Omega_k$ model, the same data combination results in $\Omega_k = 0.0021^{+0.0014}_{-0.0012}$, mildly favoring an open universe with a deviation from flatness at the $1.6\sigma$ level. We further analyze both models using the combined Pk18+DR2+PP\&SH0ES dataset. For the $\Lambda_{\rm s}$CDM$+\Omega_k$ model, we obtain a constraint of $\Omega_k = 0.0001 \pm 0.0014$, which is fully consistent with a spatially flat universe. In contrast, within the $\Lambda$CDM$+\Omega_k$ model, we find $\Omega_k = 0.0032 \pm 0.0012$, indicating a preference for an open geometry of the universe with a deviation from flatness at the $2.7\sigma$ level.  We notice that the deviation of spatial curvature from flatness remains below the $1\sigma$ level across all three data combinations for our model, indicating a clear preference for a spatially flat universe.  Meanwhile, the standard model exhibited curvature deviations in the range of approximately $1.5\sigma$ to $3\sigma$, suggesting a departure from the exact spatial flatness.\\

From both the left and right panels of Fig.\ref{fig1}, we observe that the spatial curvature parameter $\Omega_k$ is positively correlated with the $H_0$, whereas it shows a negative correlation with $S_8$ when constrained by Pk18 data alone. In the Pk18-only case, these correlations were accompanied by relatively large uncertainties, as reflected by the broad error bars. However, upon the inclusion of additional low-redshift datasets, the allowed parameter space became significantly tighter. This progressive reduction in the error bars is clearly visible in the figure, indicating that the combined datasets effectively break parameter degeneracies and lead to more robust and well-constrained estimates of $\Omega_k$, $H_0$, and $S_8$.\\

\begin{figure}[hbt!]
    \centering
    
     \includegraphics[width=0.48\linewidth]{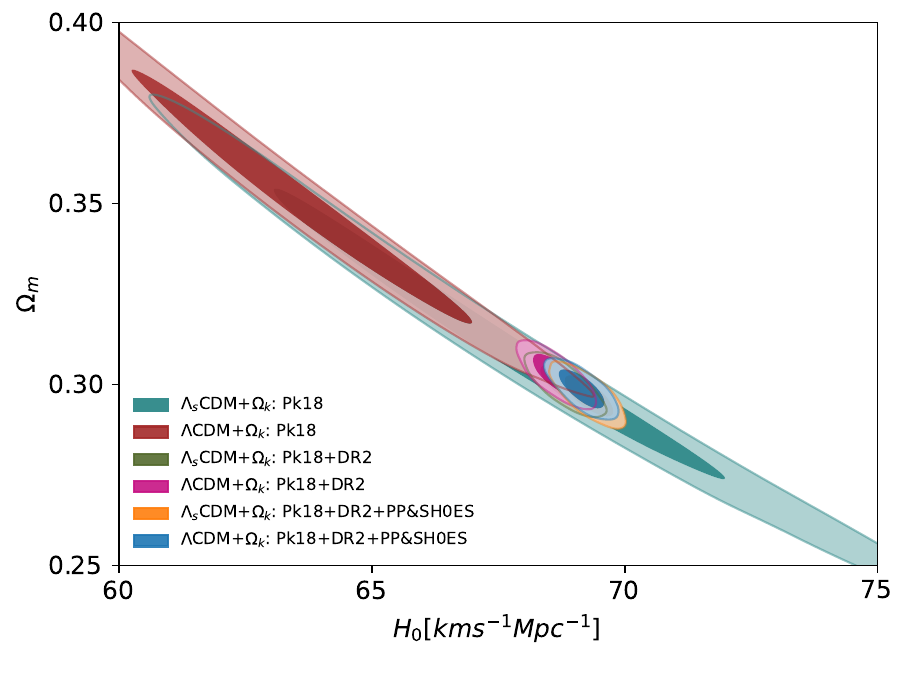}
     \includegraphics[width=0.48\linewidth]{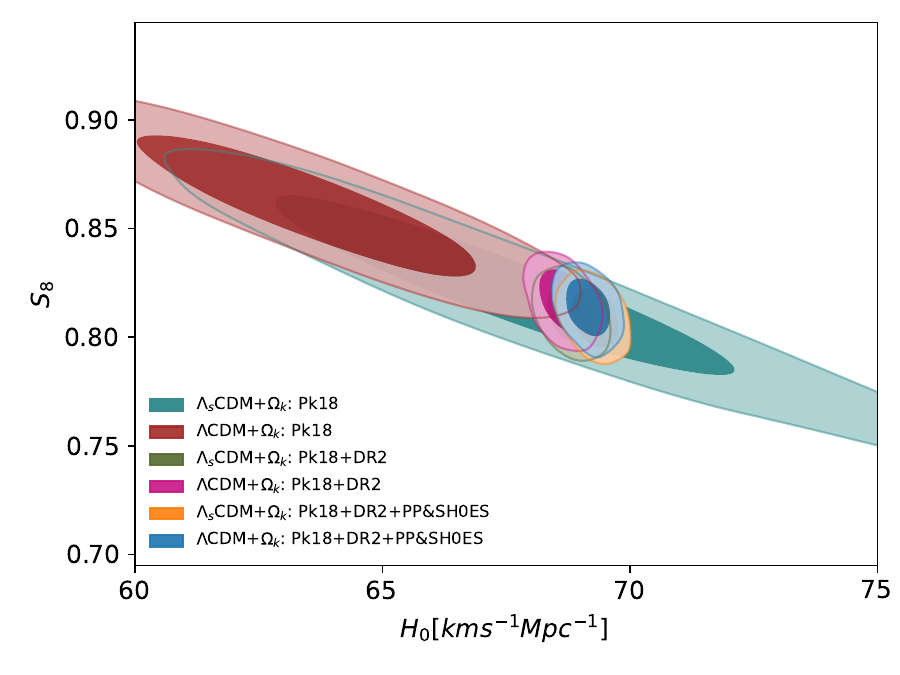}
    \caption{The 2D confidence contours (68\% and 95\% confidence levels) in the 
$\Omega_m$--$H_0$ plane (left panel) and the  $S_8$--$H_0$ plane (right panel) for the 
$\Lambda_{\rm s}$CDM and $\Lambda$CDM models, obtained from all the considered data combinations.}
    \label{fig2}
\end{figure}

The validity of the concordance model has been questioned in recent years because major tensions between parameter estimates obtained from various cosmological probes and the growing precision of cosmological observations. The most prominent of these is the Hubble tension, which is the difference between the current Hubble constant deduced from the CMB and SH0ES measurements. In particular, there is a $5-6\sigma$ discrepancy between CMB observations $H_{0}^{\text{CMB}}=67.40\pm0.50$ $\text{km} \text{s}^{-1} \text{Mpc}^{-1}$ \cite{ref1}  and SH0ES data $H_{0}^{\text{SH0ES}}=73.04\pm1.04$ $\text{km} \text{s}^{-1} \text{Mpc}^{-1}$ \cite{ref55}. As shown in Table \ref{tab1}, our model yields a constraint of $H_0 = 67.8^{+2.50}_{-3.70}$ $\text{km} \text{s}^{-1} \text{Mpc}^{-1}$ from the Pk18 dataset alone, with relatively large uncertainties, and the mean value overlaps well with the Planck determination of $H_0$. The higher value of $H_0 = 70.15 \pm 3.02~\text{km}~\text{s}^{-1}~\text{Mpc}^{-1}$ was reported within the $\Lambda_{\rm s}$CDM model without curvature \cite{ref59a}. In contrast, the comparatively lower value of $H_0$ in our analysis arises from the inclusion of spatial curvature, due to the positive degeneracy between $H_0$ and $\Omega_k$. In our analysis, the CMB data allow a mild departure of $\Omega_k$ from zero at the $\sim 1\sigma$ level, which shifts the preferred parameter space toward lower values of $H_0$, thereby accounting for the observed difference. We note that the flat $\Lambda_{\rm s}$CDM model yields a slightly higher preferred value of $H_0$ rather than the $\Lambda_{\rm s}$CDM+$\Omega_k$ extension, although both results remain statistically consistent within their respective uncertainties.\\

 Upon extending the analysis to the Pk18+DR2 and Pk18+DR2+PP\&SH0ES dataset combinations, both our model and the standard model provide nearly identical best-fit values, $H_0 = 68.81\pm 0.32(68.65\pm 0.32)$ $\text{km} \text{s}^{-1} \text{Mpc}^{-1}$ for $\Lambda_{\rm s}$CDM+$\Omega_k$ ($\Lambda$CDM+$\Omega_k$) based on the Pk18+DR2 data combination. The last column of Table \ref{tab1} further confirms that the $H_0$ estimates from Pk18+DR2+PP\&SH0ES remain consistent across both models, indicating that our model favors the Planck calibration. We further compare the constraints on $H_0$ obtained from the $\Lambda_{\rm s}$CDM+$\Omega_k$ model and the standard $\Lambda_{\rm s}$CDM model \cite{ref59a} from remaining dataset combinations.. We notice that both models yield nearly identical estimates of the Hubble constant within the quoted uncertainties. This agreement arises because the curvature parameter in our analysis is weakly constrained by given data combinations, remaining below the $1\sigma$ confidence level. As a result, the inclusion of spatial curvature does not introduce any significant shift in the cosmological parameter space, particularly for $H_0$, leading to nearly indistinguishable constraints between the two models. This indicates that, for the considered dataset combinations, the impact of curvature is statistically subdominant in determining the expansion rate. In the left panel of Fig.\ref{fig2}, we present a comparison of the joint $H_0$–$\Omega_m$ confidence contours for our model and the standard model with different data combinations. Both models display a pronounced negative correlation and a strong shrink contour between $H_0$ and $\Omega_m$. The right panel of Fig.\ref{fig2} shows the $H_0$–$S_8$ plane, where the contours are also well tightened, and the constraints for both models.\\

\begin{table*}[hbt!]
	\renewcommand{\arraystretch}{1.3} 
	\caption{Marginalized constraints (at 68\% and 95\% CL with uncertainties ) on the cosmology parameters of the $\Lambda_{\rm s}$CDM+$\Omega_k$ and  $\Lambda$CDM+$\Omega_k$ models for different dataset combinations.}
	\label{tab1}
	
	\centering
	\scalebox{0.95}{
		\begin{tabular}{lccc}
			\toprule
			\textbf{Dataset}\;\;\;\;\;\;\;\; & \;\;\;\;\;\;\;\;\textbf{Pk18}\;\;\;\;\;\;\;\; &\;\;\;\;\;\;\;\; \textbf{Pk18+DR2}\;\;\;\;\;\;\;\;& \textbf{Pk18+DR2+PP\&SH0ES} \\
			\hline
			\hline
			
			\textbf{Model} & $\bm{\Lambda}_{\rm s}$CDM+$\Omega_k$ & $\bm{\Lambda}_{\rm s}$CDM+$\Omega_k$ & $\bm{\Lambda}_{\rm s}$CDM+$\Omega_k$ \\
			& \textcolor{blue}{$\bm{\Lambda}$CDM+$\Omega_k$} & \textcolor{blue}{$\bm{\Lambda}$CDM+$\Omega_k$} & \textcolor{blue}{$\bm{\Lambda}$CDM+$\Omega_k$} \\
			
			\hline
			\hline
			
			$10^2\omega_b$ & $2.250\pm 0.016$ & $2.246\pm 0.014 $ & $2.247\pm 0.015$ \\
			& \textcolor{blue}{$2.248\pm0.016$} & \textcolor{blue}{$2.242\pm 0.015$} & \textcolor{blue}{$2.245\pm 0.015$} \\
			
			$\omega_{\rm cdm}$ & $0.1184\pm 0.0015$ & $0.1189\pm 0.0012$ & $0.1194\pm 0.0013$ \\
			& \textcolor{blue}{$0.1187\pm 0.0014$} & \textcolor{blue}{$0.1196\pm 0.0013$} & \textcolor{blue}{$0.1198\pm 0.0012$} \\
			
			$100\theta_s$ & $1.04201\pm 0.00030$ & $1.04195\pm 0.00030$ & $1.04203 \pm 0.00028$ \\
			& \textcolor{blue}{$1.04194\pm 0.00030$} & \textcolor{blue}{$1.04192\pm 0.00030$} & \textcolor{blue}{$1.04191\pm 0.00029$} \\
			
			$\ln(10^{10}A_s)$ & $3.034^{+0.012}_{-0.016}$ & $3.042^{+0.012}_{-0.014}$ & $3.046 \pm 0.013$ \\
			& \textcolor{blue}{$3.028\pm 0.017$} & \textcolor{blue}{$3.051\pm 0.014$} & \textcolor{blue}{$3.053^{+0.013}_{-0.016}$} \\
			
			$n_s$ & $0.9697\pm 0.0047$ & $0.9669\pm 0.0042 $ & $0.9689 \pm 0.0034$ \\
			& \textcolor{blue}{$0.9686^{+0.0042}_{-0.0046}$} & \textcolor{blue}{$0.9663^{+0.0040}_{-0.0045}$} & \textcolor{blue}{$0.9656\pm 0.0041$} \\
			
			$\tau_{\rm reio}$ & $0.0515^{+0.0051}_{-0.0078}$ & $0.0539\pm 0.0066$ & $0.0538^{+0.0061}_{-0.0073}$ \\
			& \textcolor{blue}{$0.0481\pm 0.0079 $} & \textcolor{blue}{$0.0573^{+0.0067}_{-0.0076}$} & \textcolor{blue}{$0.0579^{+0.0065}_{-0.0079}$} \\
			
			$z_\dagger$ & $> 1.60$ (95\% CL)& $> 2.33$ (95\% CL) & $> 2.35$ (95\% CL) \\
			& \textcolor{blue}{$-$} & \textcolor{blue}{$-$} & \textcolor{blue}{$-$} \\

			$H_0$ [km/s/Mpc] & $67.8^{+2.50}_{-3.70}$ & $68.81\pm 0.32$ & $69.28\pm 0.31$ \\
			& \textcolor{blue}{$63.9\pm 2.10$} & \textcolor{blue}{$68.65\pm 0.32$} & \textcolor{blue}{$69.15\pm 0.30$} \\
			
			$M_B$ [mag] & $-$ & $-$ & $-19.384^{+0.018}_{-0.019}$ \\
			& \textcolor{blue}{$-$} & \textcolor{blue}{$-$} & \textcolor{blue}{$-19.386\pm 0.009 $} \\
			
			$\Omega_m$ & $0.310^{+0.031}_{-0.026}$ & $0.300\pm 0.0037 $ & $0.297\pm 0.0036$ \\
			& \textcolor{blue}{$0.348\pm 0.021$} & \textcolor{blue}{$0.303\pm 0.0039$} & \textcolor{blue}{$0.299\pm 0.0035$} \\
			
			$\Omega_k$ & $-0.0059^{+0.0064}_{-0.0054}$ & $-0.0010\pm 0.0014$ & $0.0001\pm 0.0014 $ \\
			& \textcolor{blue}{$-0.0095^{+0.0064}_{-0.0056}$} & \textcolor{blue}{$0.0021^{+0.0014}_{-0.0012}$} & \textcolor{blue}{$0.0032\pm 0.0012$} \\
			
			$\sigma_8$ & $0.808^{+0.010}_{-0.014}$ & $0.811\pm 0.0062$ & $0.813\pm 0.0062$ \\
			& \textcolor{blue}{$0.796\pm 0.011 $} & \textcolor{blue}{$0.813\pm 0.0066$} & \textcolor{blue}{$0.815\pm 0.0064$} \\
			
			$S_8$ & $0.820^{+0.032}_{-0.024}$ & $0.811\pm 0.0089 $ & $0.809\pm 0.0089$ \\
			& \textcolor{blue}{$0.856\pm 0.019$} & \textcolor{blue}{$0.817\pm 0.0094 $} & \textcolor{blue}{$0.814\pm 0.0087$} \\
			
			$t_0$ [Gyr] & $13.91^{+0.31}_{-0.28}$ & $13.74^{+0.047}_{-0.053}$ & $13.68\pm 0.049$ \\
			& \textcolor{blue}{$14.19\pm 0.240$} & \textcolor{blue}{$13.685\pm 0.047$} & \textcolor{blue}{$13.629\pm 0.044$} \\

			$\chi^2_{\rm min}$ & $2778.20$ & $2788.56$ & $4104.70$ \\
			& \textcolor{blue}{$2775.00$} & \textcolor{blue}{$2790.94$} & \textcolor{blue}{$4108.46$} \\
			
	    	$\ln \mathcal{Z}$ & $-1426.84$ & $-1435.24$ & $-2094.17$ \\
	    	& \textcolor{blue}{$-1425.77$} & \textcolor{blue}{$-1436.39$} & \textcolor{blue}{$-2095.69$} \\
		
		   $\ln \mathcal{B}_{ij}$ & $-1.06$ & $1.15$ & $1.52$ \vspace{0.2cm}\\
			
			$\Delta{\rm AIC}$ & $5.20$ & $0.38$ & $1.76$ \\

			\hline
			\hline
		\end{tabular}
	}
	
\end{table*}

Another tension that has attracted considerable attention in recent years is the $S_8$ tension within the standard model, which is associated with measurements of the amplitude of matter clustering. It is quantified by $S_8 \equiv \sigma_8 \sqrt{\Omega_m/0.3}$, $\Omega_m$ with the fluctuation amplitude $\sigma_8$ on $8\,h^{-1}\,\mathrm{Mpc}$ scales. Assuming the $\Lambda$CDM framework, Planck CMB observations predict a higher matter clustering amplitude of $S_8 \approx 0.83$ \cite{ref1}, while weak lensing surveys including KiDS, DES-Y3, and HSC prefer lower values near $S_8 \approx 0.76$ \cite{ref60,ref61,ref62}, resulting in a $\sim 2\!-\!3\sigma$ tension. This discrepancy was partially mitigated when DES-Y3 and KiDS-1000 data were combined, yielding a joint constraint of $S_8 \approx 0.79$ \cite{ref63}. Recently,  KiDS-Legacy analysis reported $S_8 = 0.815^{+0.016}_{-0.021}$ \cite{ref64}, showing good consistency with Planck and reducing the tension to $0.73\sigma$. In this work, we obtained a constraint of $S_8=0.820^{-0.024}_{+0.032}$ for our model from the Pk18 dataset alone, {which shows good agreement with the mean value of the KiDS-Legacy result; however, the associated $1\sigma$ uncertainty is slightly large. In contrast, the  $\Lambda$CDM model yields a higher mean value of $S_8 = 0.856 \pm 0.019$ from the same data. Upon incorporating the DR2 dataset in combination with Pk18, both models yield similar mean values of $S_8 \approx 0.811$, which are in closer agreement with the KiDS-Legacy measurement at $1\sigma$ error bar. A consistent trend is observed for the full dataset combination Pk18+DR2+PP\&SH0ES, where both models again found nearly identical constraints on $S_8$, as summarized in Table \ref{tab1}. Overall, the $S_8$ constraints derived for both models using all dataset combinations are compatible with the KiDS-Legacy results at the $1\sigma$ level} \footnote{We note that the $S_8$ values reported by DES and KiDS are derived under the assumption of a flat $\Lambda$CDM cosmology. Since our analysis is based on a non-standard cosmological model, a direct comparison should be interpreted with caution due to possible model dependence in the inferred $S_8$ values.}\cite{ref64}.\\

 Furthermore, relatively mild tension has recently emerged between the matter density parameter $\Omega_m$ inferred from uncalibrated supernova luminosity distance–redshift relations and that obtained from uncalibrated  BAO distance measurements. The Pantheon+ (PP) supernova dataset yields $\Omega_m = 0.334 \pm 0.018$, which is in good agreement with the DES-Y5 and Union3 supernova samples \cite{ref65,ref66} , whereas recent BAO measurements from  DESI \cite{ref53} favor a lower value of $\Omega_m = 0.295 \pm 0.015$. Although the discrepancy is only at the $\sim 2\sigma$ level, it is particularly noteworthy as it is independent of the calibration of both SN absolute magnitudes and the BAO sound horizon. In this work, we consider two combined datasets, namely Pk18+DR2 and Pk18+DR2+PP\&SH0ES, as summarized in Table \ref{tab1}. For these combinations, the inferred constraint on $\Omega_m$, for our model lies close to $\sim 0.30$, which is in good agreement with the mean value reported by the DESI collaboration and aligns with the same constraint on $\Omega_m$ at the $1\sigma$ level reported by the $\Lambda_{\rm s}$CDM model without curvature \cite{ref59a}. Therefore, the effect of curvature is negligible within the sign-switch cosmology model. However, when using the Pk18 dataset alone, the mean value of $\Omega_m$ obtained for our model shows a slight deviation from the DESI-calibrated result \cite{ref53}, indicating a mild tension in this specific case.\\

 \begin{figure}[hbt!]
 	\centering
 	\includegraphics[width=1.\linewidth]{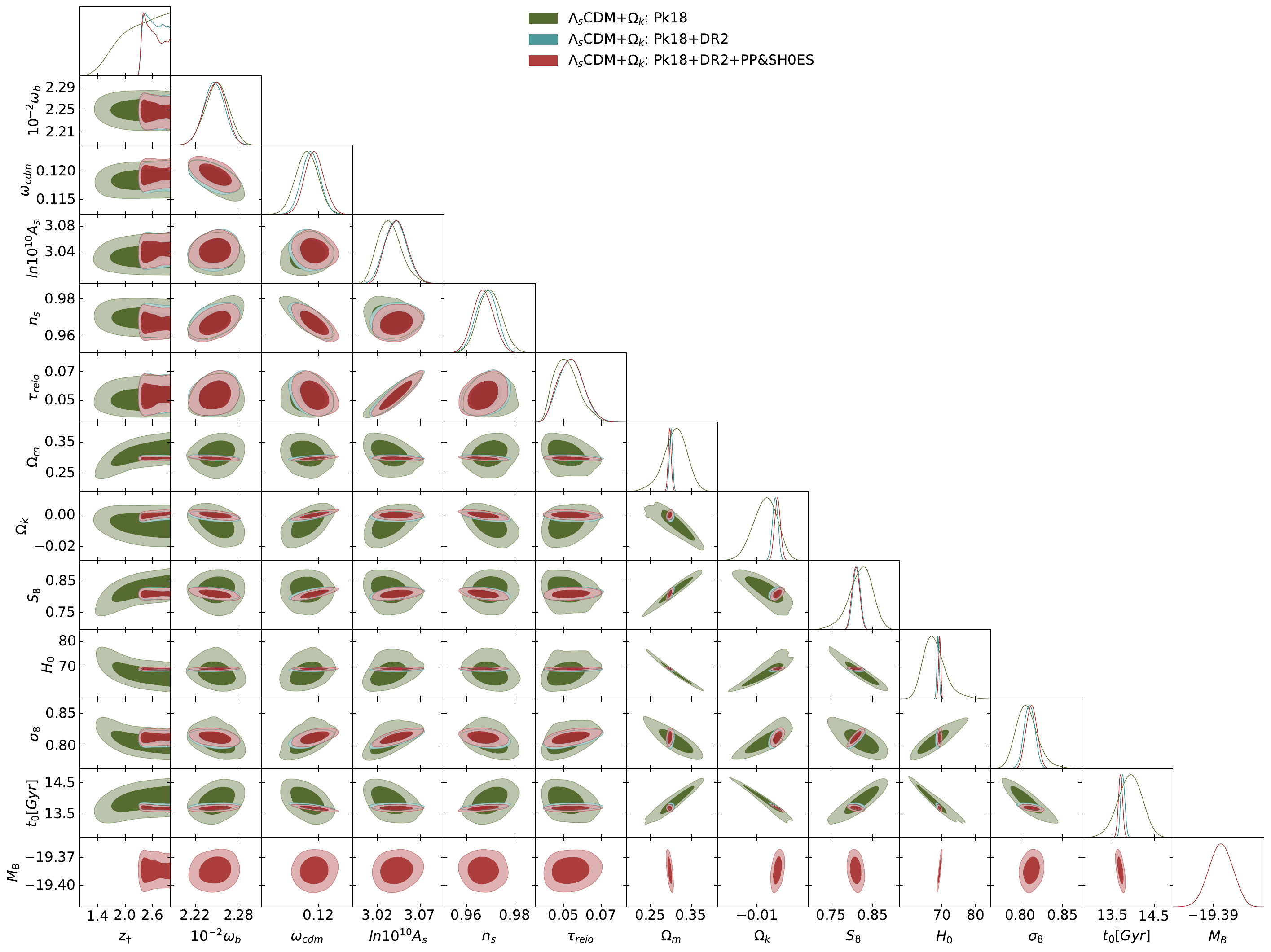}
 	\caption{The joint $68\%$ and $95\%$ confidence level contours alongside the 1-D and 2-D marginalized posterior distributions for the cosmological parameters of the modified $\Lambda_s\text{CDM}+\Omega_k$ model. The constraints are obtained using three observational data combinations: $\text{Pk18}$ , $\text{Pk18+DR2}$ , and $\text{Pk18+DR2+PP\&SH0ES}$.}
 	\label{fig3}
 \end{figure}

In our model, the transition redshift $z_\dagger$ originates from the graduated dark energy framework and characterizes a sign-switching cosmological constant from negative ($z > z_\dagger$) to positive ($z < z_\dagger$) \cite{ref46}. For large negative values of $\lambda$ ($\rho_{\rm inert} \propto \rho^\lambda < 0$),  the transition becomes sharp, and the model effectively behaves as a sign-switching cosmological constant, with $z_\dagger \sim 2.3$  preferred by observations. The prior range for $z_\dagger \in [1,3]$ was first introduced in the study by Akarsu et al.\cite{ref47}. In this study, we discuss an additional free parameter of our model, $z_{\dagger}$, whose impact on the spatial curvature parameter was investigated. As shown in the one–dimensional marginalized distributions of the triangle plot in Fig.\ref{fig3}, $z_{\dagger}$ remained largely unconstrained when only the Pk18 dataset was used. However, when the other two data combinations were considered, a clear lower bound and sharper peak emerged. The obtained constraints on $z_{\dagger}$ are broadly consistent with those reported in previous studies, indicating that spatial curvature does not exert a strong influence on this parameter. In Fig.\ref{fig4}, all the baseline and derived parameters within the $\Lambda$CDM$+\Omega_k$ model are well constrained, and their corresponding numerical values are presented in Table \ref{tab1} for three different combinations of data sets.\\

\begin{figure}[hbt!]
	\centering
	\includegraphics[width=1\linewidth]{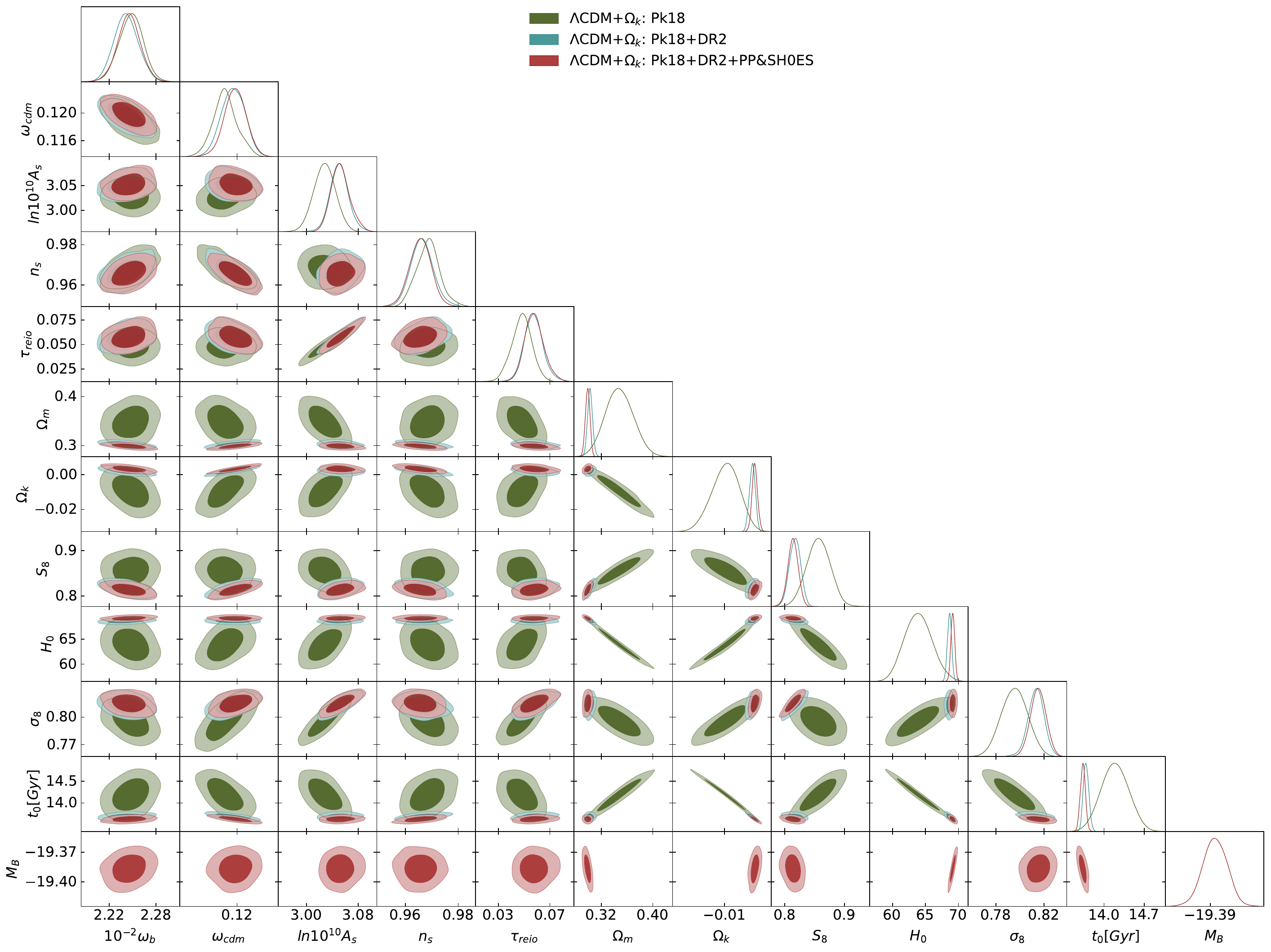}
	\caption{This triangular plot corresponds to the 1-D and 2-D posterior distributions and joint contours, respectively, for the free and derived parameters of $\Lambda$CDM+$\Omega_k$  constrained by three different combinations of observational data sets.}
	\label{fig4}
\end{figure}

Finally, we performed the Akaike Information Criterion (AIC)~\cite{ref67} to compare the {$\Lambda_{\rm s}$CDM$+\Omega_k$} models based on their fit to different combinations of observational data such as PK18, Pk18+DR2, and Pk18+DR2+PP\&SH0ES, where $\mathrm{AIC} = \chi^2_{\text{min}} + 2M$, and $M$ is the number of model-independent free parameters considered. To clearly illustrate the differences in AIC values between the {$\Lambda$CDM$+\Omega_k$} and {$\Lambda_{\rm s}$CDM$+\Omega_k$} models, we adopted the $\Lambda$CDM model as the reference with the smallest AIC value, then we set $\Delta \rm{AIC}_{{\Lambda}\rm{CDM}}$ to  zero. The resulting values of $\Delta \mathrm{AIC}$ are presented in the last row of Table \ref{tab1}, where $\Delta \mathrm{AIC} = \mathrm{AIC}_{{\Lambda_s}\mathrm{CDM}+\Omega_k} - \mathrm{AIC}_{\Lambda \mathrm{CDM}+\Omega_k}$. Thus, $\Delta\mathrm{AIC}>0$ indicates a preference for the standard $\Lambda\mathrm{CDM}+\Omega_k$ model, whereas $\Delta\mathrm{AIC}<0$ favors the $\Lambda_{\rm s}\mathrm{CDM}+\Omega_k$ model.\\

In  Table \ref{tab1}, it is clearly observed that when using the Pk18 data only, then $\Delta{\rm AIC} (5.20)$, this value lies in the region ($4 \leq \rm \Delta AIC  \leq 7$), indicating moderate evidence in favor of standard model relative to the our model. However, when additional datasets, such as DR2 and PP\&SH0ES, are included—specifically in the Pk18+DR2 and Pk18+DR2+PP\&SH0ES combinations—the level of agreement between our model and the standard model becomes statistically indistinguishable, as indicated by the small values of $\Delta \text{AIC}$ (see Table \ref{tab1}). This suggests that the $\Lambda_s$CDM +$\Omega_k$ model provides a competitive fit to the data while naturally preserving spatial flatness, whereas $\Lambda$CDM prefers non-zero curvature. This enhanced consistency demonstrates that our model performs effectively with a combination of astrophysical cosmological observations data and outperforms the standard $\Lambda$CDM+$\Omega_k$ model in explaining the current data trends.\\ 

Also, we present a discussion of the individual $\chi^2$ and AIC contributions from the DR2 and PP\&SH0ES datasets in order to assess the relative performance of the two models more transparently. For the DR2 dataset, we obtain $\chi^2 = 4.95$ for $\Lambda_{\rm s}$CDM$+\Omega_k$ and $\chi^2 = 4.99$ for $\Lambda$CDM$+\Omega_k$. Similarly, for the PP\&SH0ES dataset, we find $\chi^2 = 642.59$ for $\Lambda_{\rm s}$CDM$+\Omega_k$ and $\chi^2 = 642.98$ for $\Lambda$CDM$+\Omega_k$. These results show that the fit quality of the two models is very similar for each dataset individually. Moreover, the corresponding $\Delta$AIC difference between the models remains below 2, which indicates that neither model is statistically preferred over the other and that both provide comparably competitive fits to the individual datasets.\\

We evaluate the performance of the $\Lambda_{\rm s}$CDM$+\Omega_k$ and $\Lambda$CDM$+\Omega_k$ models through Bayesian evidence \cite{ref67a}, using the publicly available \texttt{MCEvidence}\footnote{\url {https://github.com/yabebalFantaye/MCEvidence} package}. Their relative preference is quantified by the Bayes factor,

\begin{equation}
	\ln \mathcal{B}_{ij}
	=
	\ln \mathcal{Z}_{\Lambda_s{\rm CDM}+\Omega_k}
	-
	\ln \mathcal{Z}_{\Lambda{\rm CDM}+\Omega_k}.
\end{equation}
According to the Jeffreys scale, the strength of Bayesian evidence is classified as inconclusive for $0 \leq |\ln \mathcal{B}_{ij}| < 1$, weak for $1 \leq |\ln \mathcal{B}_{ij}| < 2.5$, moderate for $2.5 \leq |\ln \mathcal{B}_{ij}| < 5$, strong for $5 \leq |\ln \mathcal{B}_{ij}| < 10$, and very strong for $|\ln \mathcal{B}_{ij}| \geq 10$. The sign of $(\ln \mathcal{B}{ij})$ indicates the direction of model preference: positive values of $(\ln \mathcal{B}{ij})$ favor the $\Lambda_{\rm s}\mathrm{CDM}+\Omega_k$ model over the $\Lambda\mathrm{CDM}+\Omega_k$ model, whereas negative values favor the $\Lambda\mathrm{CDM}+\Omega_k$ model. We notice in Table~\ref{tab1} that the $(\ln \mathcal{B}_{ij})$ value represent weak evidence for the $\Lambda$CDM$+\Omega_k$ model over $\Lambda_{\rm s}$CDM$+\Omega_k$ from Pk18 data alone. However, analyses based on the Pk18+DR2 and Pk18+DR2+PP\&SH0ES combinations indicate weak evidence in favor of the $\Lambda_{\rm s}$CDM$+\Omega_k$ model.\\

A sign-switching cosmological constant introduces a fundamentally new dynamical picture of the universe in which the vacuum energy does not act as a permanently repulsive component, but instead undergoes a transition between attractive ($\Lambda < 0$) and repulsive ($\Lambda > 0$) phases. During the negative phase, the expansion rate is effectively reduced, weakening the Hubble friction term and allowing gravitational instability to dominate, which leads to an enhanced and more rapid growth of matter perturbations. This results in an early-time boost in structure formation, potentially increasing clustering efficiency and altering the buildup of large-scale structures. As the universe evolves and the cosmological constant switches sign, the vacuum energy becomes repulsive, driving accelerated expansion and suppressing further growth. This transition naturally produces a non-monotonic evolution of density perturbations, leaving characteristic imprints in observables such as the growth rate $f\sigma_8$, the matter power spectrum, and the evolution of gravitational potentials. In particular, the transition epoch acts as a dynamical turning point where both background expansion and perturbation growth undergo qualitative changes, providing a clear and testable deviation from the smooth behavior predicted by the standard $\Lambda$CDM model.  In Ref.\cite{ref68}, the evolution of matter perturbations remains largely consistent with the $\Lambda$CDM scenario, with only small deviations that lie within current observational uncertainties. This study \cite{ref68} also reports a mild enhancement in structure growth perturbations; this effect is not sufficiently large to produce a clear observational distinction from the standard model.\\

Beyond structure formation, the sign-switching scenario has deeper implications for both observations and fundamental physics. The competing effects of early-time enhancement and late-time suppression of growth offer a natural mechanism to alleviate the $\sigma_8$ tension, while the non-trivial evolution of gravitational potentials around the transition can leave detectable signatures in the late-time Integrated Sachs--Wolfe (ISW) effect in the CMB. Importantly, this framework breaks the usual degeneracy between expansion history and growth of structures: even if the background evolution $H(z)$ mimics that of $\Lambda$CDM, the perturbation sector can exhibit distinct behavior, making growth-based probes especially powerful. From a theoretical perspective, a sign-changing cosmological constant suggests that vacuum energy is not a true constant but an effective, dynamical quantity, possibly arising from modified gravity or quantum vacuum effects. Ensuring a smooth transition is crucial to maintain perturbative stability and avoid divergences, reinforcing the idea that such models must be carefully constructed. Altogether, sign-switching cosmology provides a rich and physically motivated extension of the standard paradigm, with clear observational consequences and the potential to address some of the most persistent tensions in modern cosmology.\\

Several studies of sign-switching cosmological models indicate that the transition redshift parameter $z_\dagger$ is not uniformly constrained across different dataset combinations. Using the CMB dataset alone, a lower bound on $z_\dagger$ is typically obtained, while the upper bound remains poorly constrained or effectively unbounded \cite{ref45}. A similar behavior is observed when combining CMB with full BAO data \cite{ref46}, where a lower bound exists but the upper bound remains weak. When Pantheon, Lyman-$\alpha$, and CMB data are combined, an upper bound begins to emerge; however, the constraint on $z_\dagger$ is still not fully stable and may not exhibit a smooth posterior distribution. Interestingly, the inclusion of an $M_B$ prior in the combined analysis leads to a significant improvement, yielding well-defined and smooth constraints with both lower and upper bounds on $z_\dagger$ \cite{ref47,ref48}. In contrast, analyses performed without CMB  data show that $z_\dagger$ often remains largely unconstrained, with neither bound being well established \cite{ref59a}. The behavior highlights a degeneracy between $z_\dagger$ and $H_0$, since the $M_B$ prior is directly related to $H_0$, while $H_0$ itself is correlated with the transition parameter $z_\dagger$ through the sign-switch model. Overall, these results indicate that robust constraints on $z_\dagger$ require a careful combination of datasets, particularly those that break the degeneracy between $H_0$ and $z_\dagger$. Future high-redshift observations are expected to significantly improve constraints on $z_\dagger$. In particular, Ly-$\alpha$ BAO measurements and forthcoming 21cm intensity mapping surveys will provide direct access to the high-redshift universe, thereby offering the potential to tightly constrain the transition redshift and test the underlying sign-switching dark energy scenario in greater detail.\\

The following challenges emerge in the sign-switching cosmological constant model with respect to observational data and the treatment of a rapid transition parameter $z_\dagger$.\\

\begin{enumerate}
	
\item  The $\Lambda_{\rm s}$CDM model is that its additional free parameter, $z_\dagger$, remains poorly constrained by the most fundamental cosmological datasets, including CMB, BAO, and Pantheon observations. In particular, only a lower bound on $z_\dagger$ is typically obtained, while a robust upper bound is absent. This weak constraint indicates that the data do not strongly require the sign-switching feature, thereby limiting the predictive power of the model.

\item The $\Lambda_{\rm s}$CDM model does not appear to be favored by low-redshift observational data (Supernova samples), because both the lower and upper bounds of $z_\dagger$ are weak or absent, indicating that these datasets alone are insufficient to tightly constrain the sign-switching feature.

\item The $\Lambda_{\rm s}$CDM model is that it asymptotically approaches the standard $\Lambda$CDM model as the transition redshift $z_\dagger$ shifts to higher values. In this limit, the effect of the sign-switching feature becomes negligible, and the model effectively reduces to $\Lambda$CDM. Consequently, the same underlying issues and tensions present in the standard model may reappear, thereby limiting the distinctiveness and explanatory power of the $\Lambda_s$CDM framework.

\item To avoid an unphysical discontinuity, we replace the sign function with a smooth approximation using a hyperbolic tangent, $\Lambda_s(z) = \Lambda_{\rm dS},\tanh[(z_\dagger - z)]$, thereby ensuring a continuous expansion history. However, within this minimal smoothing framework, we find that the cosmological parameters are not tightly constrained. In particular, the inferred value of $H_0$ can shift to relatively higher values (e.g., $H_0 \sim 75^{+3.4}_{-4.6}$ km s$^{-1}$ Mpc$^{-1}$ from CMB data), indicating a lack of robustness in the parameter estimation. A more general formulation, $\Lambda_{\rm s}(z) = \Lambda_{\rm dS},\tanh[\eta (z_\dagger - z)]$, introduces a parameter $\eta > 1$ that controls the sharpness of the transition \cite{ref51}. While this additional freedom can improve the fit and stabilize the inferred constraints when $\eta$ is fixed to sufficiently large values, it comes at the cost of introducing an extra degree of freedom. More importantly, obtaining well-constrained cosmological parameters within a smooth-transition framework effectively requires the inclusion of such an additional parameter, which itself constitutes a limitation of the model. Furthermore, fixing $\eta$ a priori, rather than constraining it directly from the data, reduces the predictive power and introduces an element of arbitrariness. This highlights an intrinsic limitation of the approach, namely the trade-off between achieving a better fit and maintaining a fully data-driven inference.

\item A potential limitation of the present $\Lambda_{\rm s}$CDM framework arises from the assumption of an abrupt sign transition in the cosmological constant at the transition redshift $z_{\dagger}$. Such an instantaneous change introduces a discontinuity in the cosmological dynamics, leading to the appearance of a Type-II (sudden) singularity. Although previous studies have shown that this singular behavior has a negligible impact on the formation and evolution of large-scale cosmic structures, its presence may indicate that the abrupt-transition scenario is only an effective phenomenological description rather than a complete physical model. A more realistic treatment may require a continuous transition mechanism, which can soften the discontinuity and potentially avoid the emergence of sudden singularities.

\item The present analysis considers only an abrupt sign transition in the vacuum component. A smooth-transition realization may lead to different parameter constraints and potentially alter the model's statistical performance. Since such an investigation has not been carried out here, the robustness of the obtained constraints under a smooth-transition implementation remains an open question and constitutes a limitation of the present study.

\end{enumerate}

\section{Conclusion}

 Over the last five years, the sign-switching dark energy scenario has gained considerable attention as a promising framework for alleviating the major cosmological tensions associated with the standard $\Lambda$CDM model. However, in this study, we explored the spatial curvature of the universe within the framework of the sign-switch dark energy model, denoted as the $\Lambda_{\rm s}$CDM+$\Omega_k$ scenario. The departure of this model from the standard framework is governed by a single additional parameter, $z_{\dagger}$, which represents the redshift at which the cosmological constant changes sign. As $z_{\dagger}\rightarrow\infty$, the sign-switching feature of the cosmological constant becomes irrelevant, and the model smoothly reduces to the standard $\Lambda$CDM framework. The recent observational hints of deviations from perfect flatness and persistent tensions in the standard cosmological model, we examined both the role of curvature and a time-dependent dark energy component that undergoes a sign change, offering an alternative phenomenological description of cosmic expansion.\\

Our results highlight several key aspects of cosmic geometry and dynamics, while also revealing the inherent parameter degeneracies of the framework. Using Pk18 data alone, we find a mild preference for a closed universe in both the
 $\Lambda_{\rm s}$CDM$+\Omega_k$ and
 $\Lambda$CDM$+\Omega_k$ scenarios, in agreement
 with earlier Planck-only analyses. However, the inclusion of low-redshift information from DR2 measurements and the PantheonPlus$\&$SH0ES Type$~$Ia supernova samples substantially tightens the curvature constraints and breaks the degeneracies between $\Omega_k$, $H_0$, and $S_8$. Within the $\Lambda_{\rm s}$CDM$+\Omega_k$ framework, all considered data combinations remain fully consistent with a spatially flat universe, with deviations from flatness remaining below the $1\sigma$ level. In contrast, the standard $\Lambda$CDM$+\Omega_k$ model exhibits a shifting preference toward an open geometry when additional low-redshift datasets are included, reaching up to $\sim 2.7\sigma$ for the full data combination. We also find that $\Omega_k$ is positively correlated with the Hubble constant and negatively correlated with the clustering parameter $S_8$, particularly in CMB-only analyses. These correlations are reduced when combining high- and low-redshift probes, leading to tighter localized constraints. The model provides $S_8$ estimates ($\sim 0.811$) that are statistically compatible with weak lensing surveys such as KiDS-Legacy, thereby mitigating the clustering tension observed in standard $\Lambda$CDM.\\

Also, the transition redshift $z_{\dagger}$ characterizes the sign-switching epoch of the cosmological constant, marking the transition from an Anti-de Sitter (AdS)-like phase with negative vacuum energy to a late-time de Sitter (dS)-like phase with positive vacuum energy. Our analysis shows that $z_{\dagger}$ remains largely unconstrained with Pk18 data alone. However, the inclusion of the DR2 data significantly improves the constraints, yielding a well-defined lower bound on $z_{\dagger}$ and a pronounced peak in its 1D posterior distribution at the $68\%$ confidence level. A similar behavior is observed for the Pk18+DR2+PP\&SH0ES dataset combination, which also provides a lower bound with a clear posterior peak. Despite these improvements, the upper bound on $z_{\dagger}$ remains unconstrained in both dataset combinations. These results indicate that the inclusion of spatial curvature within the sign-switching cosmological model has a weak impact on the constraints of $z_{\dagger}$ and does not significantly improve its upper bounds.\\
 
 Finally, our model comparison based on the Akaike Information Criterion (AIC) indicates that the standard $\Lambda$CDM+$\Omega_k$ model receives moderate support relative to $\Lambda_{\rm s}$CDM+$\Omega_k$ from Pk18 data alone. For the joint dataset combinations, both frameworks are found to be statistically equivalent according to the AIC, showing no significant preference for the extended model over the baseline. Another Bayesian statistical analysis shows that the $\Lambda$CDM+$\Omega_k$ model has weak support against the $\Lambda_{\rm s}$CDM+$\Omega_k$ model from the Pk18+DR2 and Pk18+DR2+PP\&SH0ES datasets. \\

 Overall, the $\Lambda_{\rm s}$CDM$+\Omega_k$ framework favors a nearly spatially flat universe and consistently yields values of $H_0$ closer to the Planck calibration, along with $S_8$ values compatible with current observations. However, the sign-switching mechanism does not provide strong evidence for alleviating the Hubble tension within the curvature framework. The statistical model-selection results further indicate that both AIC and Bayesian evidence provide only inconclusive/weak support for the $\Lambda_{\rm s}$CDM$+\Omega_k$ model against the $\Lambda$CDM$+\Omega_k$ model. These results highlight the need for a more physically motivated treatment of the sign-switching transition, which deserves further investigation. In future work, it would be worthwhile to investigate a more physically motivated smooth transition in the cosmological constant, rather than abrupt sign-switching, and to confront such a scenario with the latest and upcoming cosmological datasets. Such an analysis could provide further insights into the viability of sign-switching dark energy models and their potential to address persistent cosmological tensions.
 
 \section*{APPENDIX}

 \subsection{Stress-Energy Conservation with Discontinuous Transition}
 
 The stress-energy momentum tensor for the dark energy component is given by

 \begin{equation}
 	\label{eq9}
 	T_{\mu\nu}^{\rm DE}
 	=
 	(\rho_{\rm DE}+p_{\rm DE})u_\mu u_\nu
 	+
 	p_{\rm DE}g_{\mu\nu}.
 \end{equation}

Using Eq.(\ref{eq1}), we find energy density and pressure for sign-switching dark energy model are defined as

 \begin{equation}
 	\label{eq10}
 	\rho_{\rm DE}(z)
 	=
 	\frac{\Lambda_{\rm s0}}{8\pi G}
 	\operatorname{sgn}(z_\dagger-z),
 \end{equation}
 and
 \begin{equation}
 	\label{eq11}
 	p_{\rm DE}(z)
 	=
 	-\frac{\Lambda_{\rm s0}}{8\pi G}
 	\operatorname{sgn}(z_\dagger-z).
 \end{equation}

From Eq.(\ref{eq10}) and Eq.(\ref{eq11}), we get, 

 \begin{equation}
 	\rho_{\rm DE}+p_{\rm DE}=0.
 \end{equation}
 
 The dark energy-momentum tensor therefore reduces to
 \begin{equation}
 	T_{\mu\nu}^{\rm DE}
 	=
 	p_{\rm DE}g_{\mu\nu}
 	=
 	-\rho_{\rm DE}g_{\mu\nu}.
 \end{equation}
 
 The covariant conservation of the dark energy-momentum tensor is given by
 \begin{equation}
 \nabla^\mu T_{\mu\nu}^{\rm DE}=-g_{\mu\nu}\nabla^\mu\rho_{\rm DE}.
 \end{equation}
 
 Since $\rho_{\rm DE}$ is a scalar quantity, its covariant derivative is identical to its ordinary partial derivative:
 \begin{equation}
 	\nabla^\mu\rho_{\rm DE}=\partial^\mu\rho_{\rm DE}=\begin{cases}
 		\partial^\mu
 		\left(
 		-\dfrac{\Lambda_{\rm s0}}{8\pi G}
 		\right),
 		& z>z_\dagger,\\[10pt]
 		\partial^\mu
 		\left(
 		+\dfrac{\Lambda_{\rm s0}}{8\pi G}
 		\right),
 		& z<z_\dagger.
 	\end{cases}.
 \end{equation}

 Since $\Lambda_{\rm s0}$ and $G$ are constants, it follows that
 \begin{equation}
 	\partial^\mu
 	\left(
 	\pm\frac{\Lambda_{\rm s0}}{8\pi G}
 	\right)
 	=0.
 \end{equation}

 Consequently,

 \begin{equation}
 		\nabla^\mu T_{\mu\nu}^{\rm DE}=0,
\end{equation}

 Thus, the dark energy momentum tensor is separately conserved on each branch of the evolution, i.e., on either side of the sign-switching transition.

 \subsection{Numerical Implementation in CLASS}
 
 We implement the sign-switching dark energy model within CLASS Boltzmann code. We introduce a new transition parameter, $z_\dagger$, in the numerical implementation. The parameter  $z_\dagger$ is defined in \texttt{background.h} and read this paramter through \texttt{input.c}. The sign-switching behavior is implemented in \texttt{background.c}, where the \textit{Lambda} subsection is modified using an if–else condition according to the following prescription:

\begin{equation}
\Lambda_{\rm s}=\begin{cases}
	+\Lambda_{\rm s0},
		& if\  a>\frac{1}{1+z_\dagger},\\[10pt]
		-\Lambda_{\rm s0},
		& otherwise.
	\end{cases}.
\end{equation}

Specifically, the code assigns
 \begin{lstlisting}[language=C, escapeinside={(*@}{@*)}]
 	/* Lambda */
 	if (pba->has_lambda == _TRUE_) {
 		if (a > 1./(1.+pba->(*@$\displaystyle \ z_\dagger$@*)))
 		{
 			pvecback[pba->index_bg_rho_lambda] =
 			pba->Omega0_lambda * pow(pba->H0,2);
 			rho_tot += pvecback[pba->index_bg_rho_lambda];
 			p_tot -= pvecback[pba->index_bg_rho_lambda];
 		}
 		else
 		{
 			pvecback[pba->index_bg_rho_lambda] =
 			-pba->Omega0_lambda * pow(pba->H0,2);
 			rho_tot += pvecback[pba->index_bg_rho_lambda];
 			p_tot -= pvecback[pba->index_bg_rho_lambda];
 		}
 	}
 \end{lstlisting}

 Thus, the cosmological constant switches sign at the prescribed transition redshift $z_\dagger$. Finally, the modified CLASS code is compiled, and the parameter $z_\dagger$ is varied through MontePython during the MCMC analysis.

\section*{Declaration of competing interest}
The authors declare that they have no known competing financial
interests or personal relationships that could have influenced
the work reported in this study.

\section*{AI Disclosure}
The authors did not use generative AI or AI-assisted technologies to create the figures and images included in this present work.

\section*{Data availability}
We employed the publicly available Planck 2018 data, DR2, and PP$\&$SH0ES data presented in this study. The Planck 2018 data and  PP$\&$SH0ES  compilation are publicly available on GitHub: $https://pla.esac.esa.int/pla and$\\ $https://github.com/brinckmann/montepython_public/tree/3.6/montepython/likelihoods/, respectively$. The DR2 data are accessible from the official repository at https://data.desi.lbl.gov/doc/releases/.

\section*{acknowledgments} 
\noindent  
 The authors (A. Dixit and A. Pradhan) are thankful to IUCAA, Pune, India for providing support and facilities under the Visiting Associateship program.  M. Yadav was sponsored by a senior Research Fellowship from the Council of Scientific and Industrial Research, Government of India (CSIR/UGC Ref.\ No.\ 180010603050). The authors appreciate the Reviewers and Editor for their informative remarks, which improved the manuscript in its current form.


\begin{thebibliography}{99} 

\bibitem{ref1}
N.~A.~Aghanim \textbf{et al.} [Planck Collaboration], Planck 2018 results. VI. Cosmological parameters, Astron. Astrophys. \textbf{641}, A6 (2020).

\bibitem{ref2}
C.-G.~Park and B.~Ratra, Using the tilted flat-$\Lambda$CDM and the untilted nonflat $\Lambda$CDM inflation models to measure cosmological parameters from a compilation of observational data, Astrophys. J. \textbf{882}, 158 (2019).

\bibitem{ref3}
W.~Handley, Curvature tension: Evidence for a closed universe, Phys. Rev. D \textbf{103}, L041301 (2021).

\bibitem{ref4}
E.~Di~Valentino, A.~Melchiorri, and J.~Silk, Planck evidence for a closed Universe and a possible crisis for cosmology, Nat. Astron. \textbf{4}, 196 (2019).

\bibitem{ref4a}
A.~Dixit, M.~Yadav, A.~Pradhan, and M.~S.~Barak, Beyond $\Lambda$CDM: Exploring a dynamical cosmological constant framework consistent with late-time observations, 
Ann. Phys. \textbf{483}, 170275 (2025).

\bibitem{ref5} 
G.~Efstathiou and S.~Gratton, The evidence for a spatially flat Universe,
Mon. Not. R. Astron. Soc. \textbf{496}, L91 (2020).



\bibitem{ref6}
S.~Anselmi, M.~F.~Carney, J.~T.~Giblin, S.~Kumar, J.~B.~Mertens, M.~O'Dwyer, G.~D.~Starkman, and C.~Tian, What is flat $\Lambda$CDM, and may we choose it?
J. Cosmol. Astropart. Phys. \textbf{02} (2023) 049.


\bibitem{ref7}
J.-Q. Jiang, D. Pedrotti, S. S. da Costa, S. Vagnozzi, Nonparametric late-time expansion history reconstruction and implications for
the Hubble tension in light of recent DESI and type Ia supernovae data, Phys. Rev. D \textbf{110} 12 (2024)


\bibitem{ref8}
 T. M. C. Abbott, M. Acevedo, \textit{et al.}, The Dark Energy Survey: Cosmology Results with $\sim$ 1500 New High-redshift Type Ia Supernovae Using the Full 5 yr Data Set, Astrophys. J. Lett. 973 (1) (2024) L14.

\bibitem{ref9}
M.~Denissenya, E.~V.~Linder, and A.~Shafieloo, Cosmic curvature tested directly from observations, J. Cosmol. Astropart. Phys. \textbf{04}, 041 (2018).

\bibitem{ref10}
J.~Dossett and M.~Ishak, Spatial curvature and cosmological tests of general relativity, Phys. Rev. D \textbf{86}, 103008 (2012).


\bibitem{ref11}
C.~D.~Leonard, P.~Bull, and R.~Allison, Spatial curvature endgame: Reaching the limit of curvature determination, Phys. Rev. D \textbf{94}, 023502 (2016).


\bibitem{ref12}
I.~Ciufolini and M.~Demianski, How to measure the curvature of space-time,
Phys. Rev. D \textbf{34}, 1018 (1986).


\bibitem{ref13}
J.~J.~Wei, Model-independent curvature determination from gravitational-wave standard sirens and cosmic chronometers, Astrophys. J. \textbf{868}, 29 (2018).


\bibitem{ref14}
Y.~He, Y.~Pan, D.~P.~Shi, J.~Li, S.~Cao, and W.~Cheng, High-precision measurements of cosmic curvature from gravitational wave and cosmic chronometer observations,
Res. Astron. Astrophys. \textbf{22}, 085016 (2022).


\bibitem{ref15}
J.~Zheng, F.~Melia, and T.~J.~Zhang, A model-independent measurement of the spatial curvature using cosmic chronometers and the HII Hubble diagram, arXiv:1901.05705 [astro-ph.CO] (2019).

\bibitem{ref16}
Y.~J.~Wang, J.~Z.~Qi, B.~Wang, J.~F.~Zhang, J.~L.~Cui, and X.~Zhang,
Cosmological model-independent measurement of cosmic curvature using distance sum rule with the help of gravitational waves, Mon. Not. R. Astron. Soc. \textbf{516}, 5187 (2022).


\bibitem{ref17}
G.~Bernstein, Metric tests for curvature from weak lensing and baryon acoustic oscillations, Astrophys. J. \textbf{637}, 598 (2006).

\bibitem{ref18}
C.~Clarkson, B.~Bassett, and T.~H.-C.~Lu, A general test of the Copernican Principle, Phys. Rev. Lett. \textbf{101}, 011301 (2008).

\bibitem{ref19}
D.~Sapone, E.~Majerotto, and S.~Nesseris, Curvature versus distances: Testing the FLRW cosmology, Phys. Rev. D \textbf{90}, 023012 (2014).

\bibitem{ref20}
S.~R{\"a}s{\"a}nen, K.~Bolejko, and A.~Finoguenov, New test of the Friedmann--Lema{\^\i}tre--Robertson--Walker metric using the distance sum rule, Phys. Rev. Lett. \textbf{115}, 101301 (2015).

\bibitem{ref21}
R.-G.~Cai, Z.-K.~Guo, and T.~Yang, Null test of the cosmic curvature using $H(z)$ and supernovae data, Phys. Rev. D \textbf{93}, 043517 (2016).

\bibitem{ref22}
T.~Collett, F.~Montanari, and S.~R{\"a}s{\"a}nen, Model-independent determination of $H_0$ and $\Omega_{K0}$ from strong lensing and Type Ia supernovae, Phys. Rev. Lett. \textbf{123}, 231101 (2019).


\bibitem{ref24}
S.-S.~Du, J.-J.~Wei, Z.-Q.~You, Z.-C.~Chen, Z.-H.~Zhu, and E.-W.~Liang, Model-independent determination of $H_0$ and $\Omega_{K0}$ using time-delay galaxy lenses and gamma-ray bursts, Mon. Not. R. Astron. Soc. \textbf{521}, 4963 (2023).
\bibitem{ref25}
J.-Q.~Xia, H.~Yu, G.-J.~Wang, S.-X.~Tian, Z.-X.~Li, S.~Cao, and Z.-H.~Zhu,
Revisiting studies of the statistical property of a strong gravitational lens system and model-independent constraint on the curvature of the Universe,
Astrophys. J. \textbf{834}, 75 (2017).

\bibitem{ref26}
J.-Z.~Qi, S.~Cao, S.~Zhang, M.~Biesiada, Y.~Wu, and Z.-H.~Zhu,
The distance sum rule from strong lensing systems and quasars: Test of cosmic curvature and beyond, Mon. Not. R. Astron. Soc. \textbf{483}, 1104 (2019).

\bibitem{ref27}
T.~Liu, S.~Cao, J.~Zhang, M.~Biesiada, Y.~Liu, and Y.~Lian, Testing the cosmic curvature at high redshifts: The combination of LSST strong lensing systems and quasars as new standard candles, Mon. Not. R. Astron. Soc. \textbf{496}, 708 (2020).

\bibitem{ref28}
B.~Wang, J.-Z.~Qi, J.-F.~Zhang, and X.~Zhang, Cosmological model-independent constraints on spatial curvature from strong gravitational lensing and SN Ia observations, Astrophys. J. \textbf{898}, 100 (2020).

\bibitem{ref29}
L.~Liu, L.-J.~Hu, L.~Tang, and Y.~Wu, Constraining the spatial curvature of the local Universe with deep learning, Res. Astron. Astrophys. \textbf{23}, 125012 (2023).

\bibitem{ref30}
M.~Yadav, A.~Dixit, M.~S.~Barak, and A.~Pradhan,
Constraints on spatial curvature and dark energy dynamics in the $w$CDM model from DESI DR1 and DR2,
J. High Energy Astrophys. \textbf{50}, 100514 (2026).

\bibitem{ref31}
B.~R.~Dinda,
Cosmic curvature on large-scale structures with homogeneous dark energy,
Phys. Rev. D \textbf{111}, 103533 (2025).


\bibitem{ref32}
P.~J.~Wu and X.~Zhang, Measuring cosmic curvature with non-CMB observations,
Phys. Rev. D \textbf{112}, 063514 (2025).

\bibitem{ref33}
P.~J.~Wu, Comparison of dark energy models using late-universe observations,
Phys. Rev. D \textbf{112}, 043527 (2025).

\bibitem{ref34}
B.~R.~Dinda and R.~Maartens,
Physical versus phantom dark energy after DESI: Thawing quintessence in a curved background,
Mon. Not. R. Astron. Soc. Lett. \textbf{542}, L31 (2025).


\bibitem{ref35}
S.~F.~Chen and M.~Zaldarriaga, It's all ok: Curvature in light of BAO from DESI DR2, arXiv:2505.00659 [astro-ph.CO] (2025).




\bibitem{ref36}
S.~Bhattacharya, G.~Borghetto, A.~Malhotra, S.~Parameswaran, G.~Tasinato, and I.~Zavala, Cosmological constraints on curved quintessence, J. Cosmol. Astropart. Phys. \textbf{09}, 073 (2024).


\bibitem{ref37}
V.~Yadav, M.~Yadav, S.~K.~Yadav \textbf{et al.}, Testing spatial curvature in an anisotropic extension of $w$CDM model with low redshift data, arXiv:2405.11534 [astro-ph.CO] (2024).


\bibitem{ref38}
\"O.~Akarsu, E.~Di~Valentino, S.~Kumar, M.~\"Ozyi\u{g}it, and S.~Sharma,
Testing spatial curvature and anisotropic expansion on top of the $\Lambda$CDM model, Phys. Dark Univ. \textbf{39}, 101162 (2023).

\bibitem{ref39}
W.~Yang, W.~Giare, S.~Pan, E.~Di~Valentino, A.~Melchiorri, and J.~Silk,
Revealing the effects of curvature on the cosmological models,
Phys. Rev. D \textbf{107}, 063509 (2023).

\bibitem{ref40}
J.~Stevens, H.~Khoraminezhad, and S.~Saito,
Constraining the spatial curvature with cosmic expansion history in a cosmological model with a non-standard sound horizon,
J. Cosmol. Astropart. Phys. \textbf{07}, 046 (2023).


\bibitem{ref41}
E.~Mortsell and J.~Jonsson,
A model independent measure of the large scale curvature of the universe,
arXiv:1102.4485 [astro-ph.CO] (2011).

\bibitem{ref42}
E.~Di~Dio, F.~Montanari, A.~Raccanelli, R.~Durrer, M.~Kamionkowski, and J.~Lesgourgues, Curvature constraints from large scale structure,
J. Cosmol. Astropart. Phys. \textbf{06}, 013 (2016).


\bibitem{ref43}
J.-M.~Virey, D.~Talon-Esmieu, A.~Ealet, P.~Taxil, and A.~Tilquin,
On the determination of curvature and dynamical dark energy,
J. Cosmol. Astropart. Phys. \textbf{12}, 008 (2008).


\bibitem{ref44}
E.~Di~Valentino \textbf{et al.}, Snowmass2021 - Letter of interest cosmology intertwined IV: The age of the universe and its curvature, Astropart. Phys. \textbf{131}, 102607 (2021).

\bibitem{ref45}
M.~Yadav, A.~Dixit, A.~Pradhan, and M.~S.~Barak, Empirical validation: Investigating the $\Lambda$sCDM model with new DESI BAO observations, J. High Energy Astrophys. \textbf{100453} (2025).


\bibitem{ref46}
{\"O}.~Akarsu, \textit{et al.}, Graduated dark energy: Observational hints of a spontaneous sign switch in the cosmological constant, 
Phys. Rev. D \textbf{101}, 063528 (2020).


\bibitem{ref47}
  {\"O}. Akarsu, \textit{et al.}, Relaxing cosmological tensions with a sign switching cosmological constant, Phys. Rev. D {\bf 104}, 123512 (2021).
  
 
\bibitem{ref48}
{\"O}. Akarsu,\textit{et al.}, Relaxing cosmological tensions with a sign switching cosmological constant: Improved results with Planck, BAO, and Pantheon data, 
Phys. Rev. D \textbf{108}, 023513 (2023).

\bibitem{ref49}
{\"O}. Akarsu, \textit{et al.},
$\Lambda_{\rm s}$CDM model: A promising scenario for alleviation of cosmological tensions, arXiv:2307.10899  (2023).

\bibitem{ref50}
A. Yadav, \textit{et al}.,
$\Lambda_{\rm s}$CDM cosmology: Alleviating major cosmological tensions by predicting standard neutrino properties, JCAP \textbf{2025}, 042 (2025).

\bibitem{ref51}
{\"O}. Akarsu, \textit{et al.},  
$\Lambda_{\rm s}$CDM cosmology from a type-II minimally modified gravity, arXiv:2402.07716 [astro-ph.CO] (2024).



\bibitem{ref52}
N. Aghanim \textit{et al.}, Planck2018 results: I. Overview and the cosmological legacy of Planck, Astrophysics \textbf{641}, A1 (2020).

\bibitem{ref53}
M.~Abdul~Karim \textit{et al.},
DESI DR2 results. II. Measurements of baryon acoustic oscillations and cosmological constraints,
Phys. Rev. D \textbf{112}, 083515 (2025).

\bibitem{ref53a}
M.~Yadav, A.~Dixit, M.~S.~Barak, and A.~Pradhan, Dynamical oscillations in dark energy: Joint constraints on the $w_{\sin}$CDM model from DESI, OHD, and supernova samples, 
J. High Energy Astrophys. \textbf{51}, 100556 (2026).


\bibitem{ref54}
D. Scolnic, D. Brout, \textit{et al}., The Pantheon+ Analysis: The Full Data Set and Light-curve Release, \textit{Astrophys. J.} \textbf{938}, 113 (2022).

\bibitem{ref55}
A.G Riess \textit{et al}, A Comprehensive Measurement of the Local Value of the Hubble Constant with  Uncertainty from the Hubble Space Telescope and the SH0ES Team, Astrophys. J. Lett. \textbf{934}, L7 (2022).  

\bibitem{ref56}
T. Brinckmann, J. Lesgourgues, MontePython 3: Boosted MCMC sampler and other features, phys. Dark Universe \textbf{24}, 100260 (2019).

\bibitem{ref57}
 B. Audren, J. Lesgourgues, K. Benabed, S. Prunet, Conservative constraints on early cosmology with MONTE PYTHON, JCAP \textbf{001}, 02 (2013).

\bibitem{ref58}
A. Lewis, GetDist: a python package for analysing Monte Carlo samples,  	arXiv:1910.13970 [astro-ph.IM] (2019).

\bibitem{ref59}
A. Gelman and D. B. Rubin, Inference from iterative simulation using multiple sequences, Stat. Sci. \textbf{ 7} 472 (1992).

\bibitem{ref59a}
B.~Ibarra-Uriondo and M.~Bouhmadi-L{\'o}pez, Sign-switching dark energy: Smooth transitions with recent \textit{DESI DR2} observations, arXiv:2602.12347 (2026).


\bibitem{ref60}
A. Amon, et al., Dark energy survey year 3 results: cosmology from cosmic shear
and robustness to data calibration. Phys. Rev. D 105, 023514 2022.

\bibitem{ref61}
M.Asgari, et al., Kids-1000 cosmology: cosmic shear con­
straints and comparison between two point statistics. Astron. Astrophys. 645, A104 2021.

\bibitem{ref62}
R. Dalal, et al., Hyper suprime-cam year 3 results: cosmology from cosmic shear
power spectra. Phys. Rev. D 108, 123519 2023.

\bibitem{ref63}
T.M.C. Abbott, et al., DES Y3 + KiDS-1000: consistent cosmology combining cosmic
shear surveys. Open J. Astrophys. 6. 2023 arXiv:2305.17173.

\bibitem{ref64}
A.H. Wright, et al., KiDS-Legacy: cosmological constraints from cosmic shear with
the complete Kilo-Degree Survey, Astron. Astrophys. 703, A158 2025

\bibitem{ref65}
David Rubin \textit{et al}., Union Through UNITY: Cosmology with 2,000 SNe Using a Unified Bayesian Framework,\textit{Astrophys. J.} \textbf{986}, 231 (2025).


\bibitem{ref66}
T. M. C. Abbott \textit{et al}. (DES), The Dark Energy Survey: Cosmology Results With ~1500 New High-redshift Type Ia Supernovae Using The Full 5-year Dataset, \textit{Astrophys. J. Lett.} \textbf{973}, L14 (2024)



\bibitem{ref67}
H. Akaike, A new look at the statistical model identification, IEEE transactions on automatic control \textbf{19}, 716 (1974).

\bibitem{ref67a}
R.~Trotta, Bayes in the sky: Bayesian inference and model selection in cosmology, 
Contemp. Phys. \textbf{49}, 71 (2008).


\bibitem{ref68}
M.~Bouhmadi-L{\'o}pez and B.~Ibarra-Uriondo, Cosmological perturbations for smooth sign-switching dark energy models, Phys. Dark Univ. \textbf{50}, 102129 (2025).






\end{thebibliography}
\end{document}